\documentclass[iop,onecolumn]{emulateapj}

\usepackage[dvips]{color}

\usepackage{graphics}
\usepackage{epsfig}
\usepackage{amsmath}
\usepackage{amssymb}
\newcommand\doubleprime{\mbox{$^{\prime\prime}$}}%
\newcommand{\Htwo}{\mbox{H$_{2}$}}%
\newcommand{\htwo}{H$_2$}%
\newcommand{\hst}{\textit{HST\/}}%
\newcommand{\fuse}{\emph{FUSE\/}}%
\newcommand{\cmsq}{cm$^{-2}$}%
\newcommand{\trot}{$T_{rot}$}%

\shorttitle{\htwo\ Fluorescence with Partial Frequency Redistribution}
\shortauthors{Lupu et al.}

\begin{document}

\title{OBSERVATIONS AND MODELING OF H$_{2}$ FLUORESCENCE WITH PARTIAL FREQUENCY REDISTRIBUTION IN GIANT PLANET ATMOSPHERES}

\author{Roxana E. Lupu}
\affil{Dept. of Physics and Astronomy, University of Pennsylavania,
209 S 33rd Street, Philadephia, PA 19104}
\email{lroxana@sas.upenn.edu}

\author{Paul D. Feldman, Stephan R. McCandliss}
\affil{Dept. of Physics and Astronomy, The Johns Hopkins University,
3400 N. Charles Street, Baltimore, MD 21218}

\author{Darrell F. Strobel}
\affil{Dept. of Earth and Planetary Sciences, The Johns Hopkins University,
3400 N. Charles Street, Baltimore, MD 21218}

\begin{abstract}
Partial frequency redistribution (PRD), describing the formation of the line profile, has negligible observational effects for optical depths smaller than $\sim$10$^{3}$, at the resolving power of most current instruments. However, when the spectral resolution is sufficiently high, PRD modeling becomes essential in interpreting the line shapes and determining the total line fluxes. We demonstrate the effects of PRD on the \htwo\ line profiles observed at high spectral resolution by the $Far-Ultraviolet$ $Spectroscopic$ $Explorer$ (\fuse) in the atmospheres of Jupiter and Saturn. In these spectra, the asymmetric shapes of the lines in the Lyman (v\doubleprime - 6) progression pumped by the solar Ly$\beta$ are explained by coherent scattering of the photons in the line wings. We introduce a simple computational approximation to mitigate the numerical difficulties of radiative transfer with PRD, and show that it reproduces the exact radiative transfer solution to better than 10\%. The lines predicted by our radiative transfer model with PRD, including the \htwo\ density and temperature distribution as a function of height in the atmosphere, are in agreement with the line profiles observed by $FUSE$. We discuss the observational consequences of PRD, and show that this computational method also allows us to include PRD in modeling the continuum pumped \htwo\ fluorescence, treating about 4000 lines simultaneously. 

\end{abstract}


\section{INTRODUCTION}

The study of the ultraviolet spectra of giant planets started with the $Voyager$ fly-bys and rocket observations \citep{Broadfoot:1977,Moos:1969}, and continued with Galileo and Cassini, as well as with observations from space-based observatories such as the International Ultraviolet Explorer (IUE), the Hopkins Ultraviolet Telescope (HUT), and the $Far$-$Ultraviolet$ $Spectroscopic$ $Explorer$ ($FUSE$). The increasing spectral resolution of the cited observations allowed for a better separation of the \htwo\ excitation processes, constraining the thermal structure of the planetary atmospheres. The far-UV spectra of giant planet atmospheres are dominated by molecular hydrogen emission produced by electron-impact excitation \citep{LiuDalgarno:1996,Wolven:1998}, and, to a lesser extent, by fluorescent pumping \citep{Feldman:1993}. The molecular hydrogen transitions observed in the far-UV occur via de-excitation from the excited electronic states, with most of the contributions coming from the states B$^{1}\Sigma_{u}^{+}$, C$^{1}\Pi_{u}$, EF$^{1}\Sigma_{g}^{+}$, B$'^{1}\Sigma_{u}^{+}$, D$^{1}\Pi_{u}$, B\arcsec$^{1}\Sigma_{u}^{+}$, and D$'^{1}\Pi_{u}$. Since these electronic states lie at 11 eV or more above the ground electronic state (X$^{1}\Sigma_{g}^{+}$), only photoelectrons, auroral electrons, and direct solar UV radiation have enough energy to excite these transitions. The solar far-UV continuum is however too weak to make a significant contribution to the observed spectrum, and the only observed radiatively-excited features arise from selective pumping of \htwo\ transitions coincident with strong solar emission lines. At \Htwo\ rotational temperatures below 1000~K that characterize the atmospheres of giant planets \citep{LiuDalgarno:1996,Wolven:1997}, the main mechanism generating \Htwo\ fluorescent lines is pumping by solar Lyman-$\beta$, which generates lines in the Lyman system (B$^{1}\Sigma_{u}^{+}$-X$^{1}\Sigma_{g}^{+}$). Several of these lines were observed in the HUT observations of the Jovian dayglow by \citet{Feldman:1993}. In addition, weak fluorescent emission in these lines has been used to detect \Htwo\ for the first time in the atmosphere of Mars \citep{Krasnopolsky:2001} and in a cometary coma \citep{Feldman:2002}.

We report $FUSE$ observations of Jupiter and Saturn, in the spectral range 905--1187 \AA\ at a spectral resolution of $\sim$0.08-0.12 \AA, sufficient to resolve the line profiles of the \Htwo\ fluorescent lines pumped by solar Ly$\beta$. For both planets, we observe a broad asymmetric profile for three of the Lyman (6$-$v\arcsec)P(1) lines, namely Ly(6$-$3)P(1) $\lambda$1166.8, Ly(6$-$1)P(1) $\lambda$1071.6, and Ly(6$-$2)P(1) $\lambda$1118.6, not reproduced by frequency-independent fluorescence models. These line profiles can be explained by the asymmetry in the overlap between the Ly(6$-$0)P(1) line and the solar Ly$\beta$, combined with the coherent scattering of photons in the line wings. We treat separately the electron-impact and radiative excitation contribution to the spectra, and investigate both the complete frequency redistribution (CRD) approximation, as well as the PRD solution for the formation of the Ly$\beta$-pumped \htwo\ lines. For the line-pumping model, we build synthetic spectra starting from depth dependent atmospheric models, and show that the observed line profiles cannot be reproduced under the approximation of CRD, even taking into account the spatial variation of the atmospheric composition and temperature, or the differences in the numerical integration method for the radiative transfer equations. The use of PRD on the other hand correctly describes the coherent scattering in the line wings, proven by the agreement between the model and the data, and allows an independent confirmation of the \htwo\ abundance and atmospheric structure.

The PRD effects have been thoroughly studied for the formation of resonant and subordinate lines in stellar atmospheres \citep{Shine1975,HubenyLites:1995,Uiten:2001}, but only the profiles of resonant lines with overlapping transitions have been taken into consideration for planetary atmospheres \citep{Gladstone:1982,Wallace:1989,Grif:2000,Barth:2004}. The importance of PRD has also been pointed out for the formation atomic lines in quasi-stellar objects and active galactic nuclei (AGNs) \citep{CollSouf:1986,Avrett:1988}, and for the X-ray Bowen fluorescence in AGNs \citep{Sako:2003}. The formalism presented in this paper is geared towards the treatment of cross-redistribution in fluorescent molecular lines, which has not been investigated to date. The case of molecular lines is complicated due to the multitude of transitions, the large wavelength range spanned by all possible subordinate lines, and the fine frequency grid required to accurately describe the line profiles.

The $FUSE$ observations of the atmospheres of Jupiter and Saturn represent the first detection of PRD effects in subordinate molecular fluorescent lines, and offer grounds for investigating the same process in other astrophysical environments. The far-UV \htwo\ emission can be difficult to interpret due to the contributions of different excitation mechanisms, dust extinction and scattering, as well as the large optical depth effects at high column densities. Complex models of molecular hydrogen emission in astrophysical environments have been constructed \citep[e.g., ][]{Sternberg:1989,Draine:1996,LiuDalgarno:1996,Shaw:2005}, taking into account the formation and destruction processes, the chemistry, and the detailed balance equations. However, in these models the radiative transfer of the molecular lines is often treated in a simplified manner, due to computational limitations, and does not attempt to reproduce the line profiles. Although the use of the PRD functions to characterize the emission line profiles involves adding an additional dimension to the radiative transfer problem and requires large computational resources, the consequences for the predicted spectrum can be significant.

Our approach for modeling the \htwo\ fluorescence is complementary to previous models, focusing on reproducing the line profiles, while making simplifying assumptions for the ground state population and the geometry of the system. In order to mitigate the computational difficulties, we investigate approximate solutions for the radiative transfer equation, in the case of a plane-parallel atmosphere. Starting from an optically thin approximation \citep{LiuDalgarno:1996}, we find that treating a medium of large optical depth as a series of optically thin layers reproduces the Feautrier solution for a single line at the 10\% level. This multi-layer approach is the method of choice to generate the synthetic spectra for Ly$\beta$ pumped fluorescence of \htwo, reproducing the $FUSE$ observations. This method also allows us to test the effects of PRD on continuum pumped \htwo\ fluorescence, a case involving several thousand transitions, that would otherwise be intractable. In a simple toy model for the continuum pumped \htwo\ fluorescence, using just a single layer approximation, the inclusion of PRD results in significant blending of the overlapping line wings, and has the effect of reducing the line contrast by generating a pseudo-continuum. We show that this process has important consequences for the detectability of lines for large optical depths and high spectral resolution observations.

The $FUSE$ observations are presented in Section~\ref{secobs}. Section~\ref{secmodel} describes the \htwo\ fluorescence model, the approximations used, and the introduction of PRD to reproduce the line profiles. The application to line-pumped \htwo\ in planetary atmospheres and the continuum pumping test case follow in Section~\ref{data}. A summary of our conclusions is given in Section~\ref{secsum}.

\section{$FUSE$ OBSERVATIONS OF JUPITER AND SATURN}\label{secobs}

\subsection{Observations and Data Reduction}

\begin{figure}[h]
\begin{center}
\epsscale{0.7}
\rotatebox{0}{
\plotone{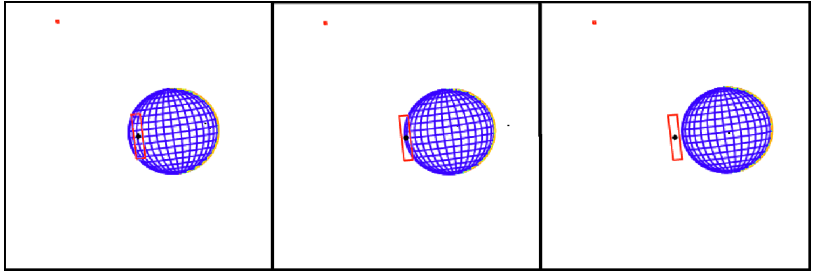}
}
\caption[Time steps for the $FUSE$ observations of Jupiter.]{Snapshot montage of the MDRS slit positions for the $FUSE$ observations of the Jupiter limb, with a time interval of 24 minutes. \label{obsjup}}
\end{center}
\end{figure}

The Jovian dayglow was recorded during drift scans of the eastern limb
of Jupiter made on 2001 October 19 and 21 using the \fuse\ MDRS ($4''
\times 20''$) aperture.  This approach was chosen to minimize the
possibility of contamination by the much stronger polar aurora and to
take advantage of possible limb brightening of \ion{D}{1} Lyman-$\beta$ emission. A single scan of $\sim$~3000~s was made during each of five consecutive orbits on both days. For each scan, the long side of the slit was aligned nearly north-south, along a chord centered on the Jovian equator. The scans started with the center of the slit positioned at 0.77~R$_{Jup}$ from the center of Jupiter, and continued towards the eastern limb a rate of $\sim$~0\farcs23~minute$^{-1}$. The leading edge of the slit crossed the limb of the planet $\sim$~700~s after the start of the scan, and $\sim$~300 s later the spacecraft crossed the terminator into orbit day. A montage of slit positions during these observations is shown in Figure~\ref{obsjup}. At the time of the observations, Jupiter's heliocentric distance was 5.14 AU, and its geocentric distance was 4.86 and 4.83~AU for the two days, respectively. 

\fuse\ consists of four separate telescope/spectrograph assemblies, two employing lithium fluoride coatings (LiF) and two using silicon carbide (SiC). Each of four separate detectors (denoted 1a, 1b, 2a, and 2b) records two spectra, one each from a LiF and a SiC channel spanning a given wavelength interval, giving a total of eight separate spectra. Details of the \fuse\ instrument have been given by \citet{Moos:2000} and \citet{Sahnow:2000}. For the uniformly filled MDRS aperture, $FUSE$ achieved a spectral resolution of 0.08 \AA. Due to thermal misalignments between channels, the Jovian spectrum was not recorded in all four channels. Only the LiF1 channel, coupled to the Fine Error Sensor used for guiding and tracking, shows the expected level of disk emission, decreasing over time as the slit drifted off the limb. The disk emission recorded by the SiC1 channel has a significantly lower signal/noise, as the effective area of this channel is a factor of 4 lower than that of LiF1. The LiF2 and SiC2 channels were off the disk for the entire time,
showing only emission from the Io plasma torus. The data were
reprocessed with the final \fuse\ pipeline, CalFUSE version 3.2.  For
each day, the data from the first half of each orbit are co-added
giving an integration time of 6330 and 7336 seconds, respectively, for
the two days, and the extracted fluxes were converted to average
brightness (in rayleighs) in the $4'' \times 20''$ aperture.

The observations of Saturn were made during 5 contiguous orbits starting at 3:20:06 UT on 2001 January 15. Each 2000~s exposure was performed using the $30'' \times 30''$ slit (LWRS) centered on the planet. At the time of the observation, Saturn's heliocentric distance was 9.11~AU, and its geocentric distance was 8.59~AU. The integrated emission from the entire Saturn disk was recorded in each observation, with an effective spectral resolution of 0.12~\AA. The data were also reprocessed with CalFUSE version 3.2, and all five orbits of data were co-added together and rebinned to 0.053~\AA\ bins. These observations have also been analyzed by \citet{Gustin2009}, who showed that the southern polar aurora contributed most of the \htwo\ emission through electron impact excitation, using archival \hst/STIS FUV-MAMA images and spectra taken on 2000 December 7--8. This conclusion is also supported by the narrow line widths, $\sim$0.10 \AA, considerably narrower than if the $19.\!''3$ Saturnian disk was the primary source of the emission in the $30'' \times 30''$ slit. The sub-Earth latitude of Saturn was $-$27.6\degr\ so that the southern polar region of Saturn was clearly visible. In contrast, both fluorescent and electron impact sources fill the entire aperture in the Jupiter observations.

\begin{figure}[tbh]
\begin{minipage}{0.48\textwidth}
\begin{center}
\epsscale{1.05}
\rotatebox{0}{
\plotone{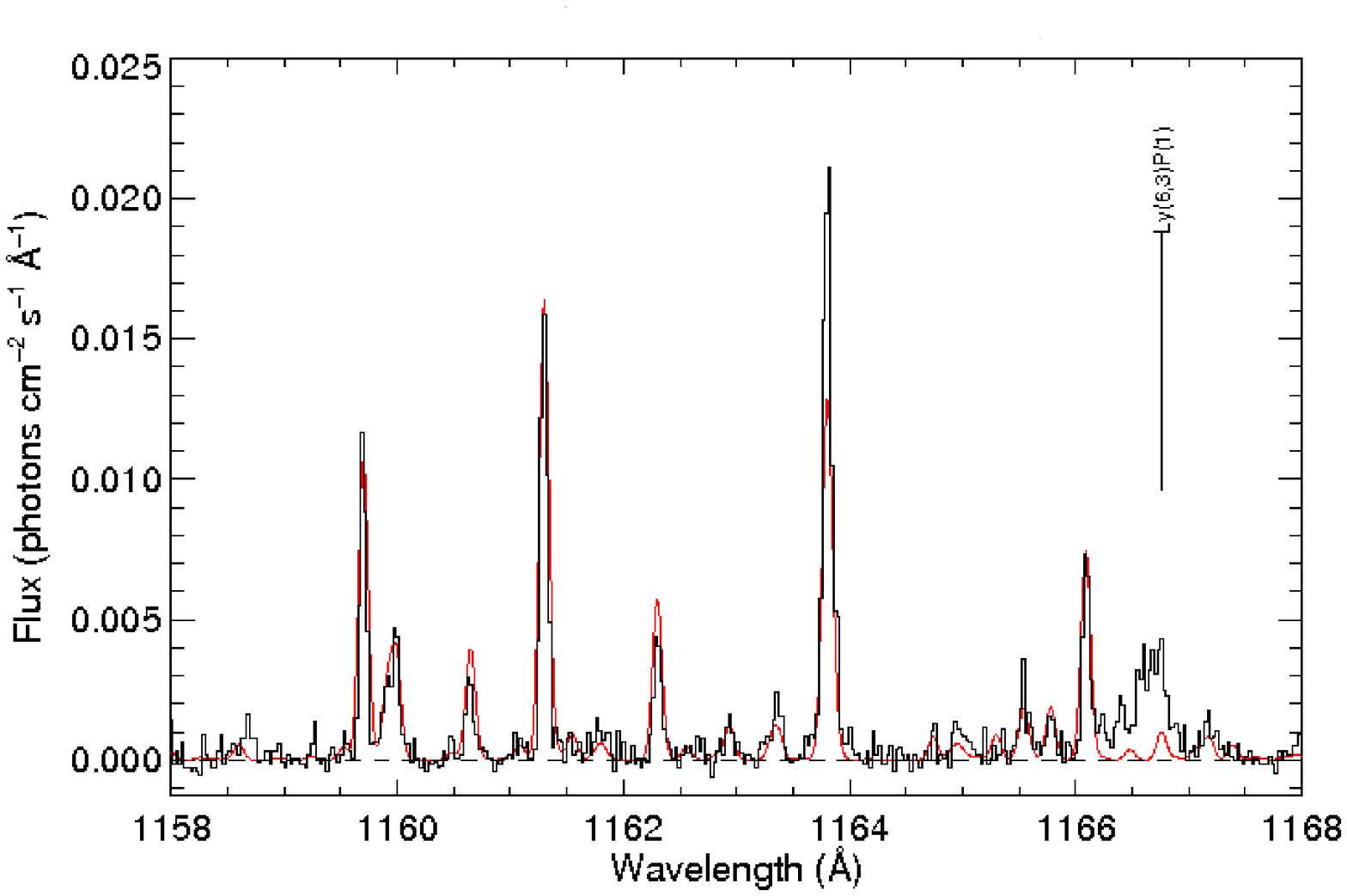}
}
\caption[$FUSE$ Spectrum of the Jovian Limb]{Selected region from the $FUSE$ spectrum of the Jovian limb, with the photoelectron impact excitation model \citep{Wolven:1998} overplotted in red. \Htwo\ fluorescence pumped by solar \ion{O}{6} $\lambda$1031.91 and Ly$\beta$ produces the excess brightness at 1163.8~\AA, and the feature at 1166.76~\AA\ (labeled), respectively. \label{fusejup}}
\end{center}
\end{minipage}\hfill
\begin{minipage}{0.48\textwidth}
\begin{center}
\epsscale{1.05}
\rotatebox{0}{
\plotone{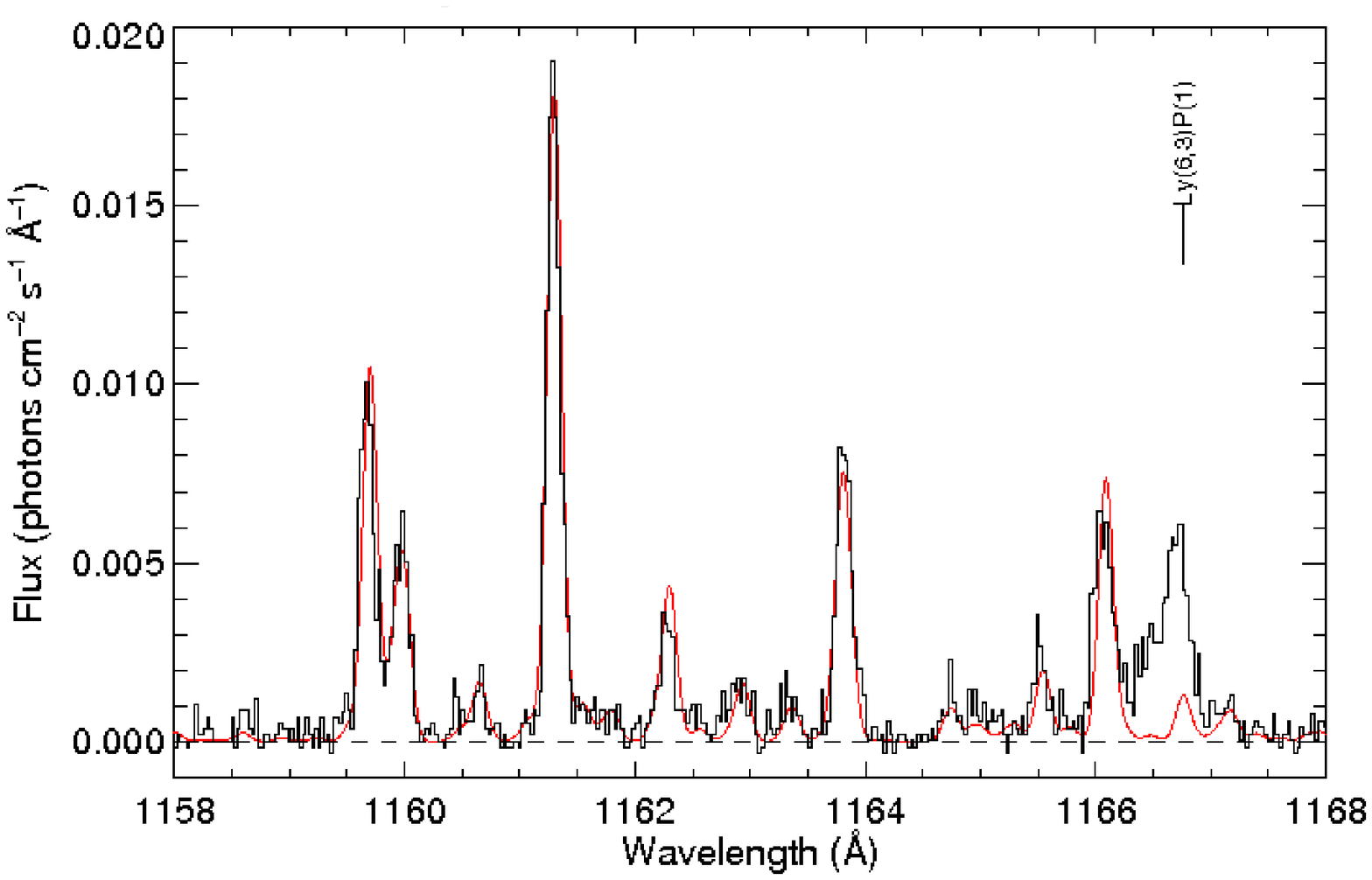}
}
\caption[$FUSE$ Spectrum of Saturn]{Same as Figure~\ref{fusejup}, for the Saturn disk. The red model shows the \Htwo\ emission generated by collisional excitation with auroral electrons. \label{specsat}}
\end{center}
\end{minipage}
\end{figure} 

\begin{figure}
\begin{center}
\epsscale{0.40}
\rotatebox{90}{
\plotone{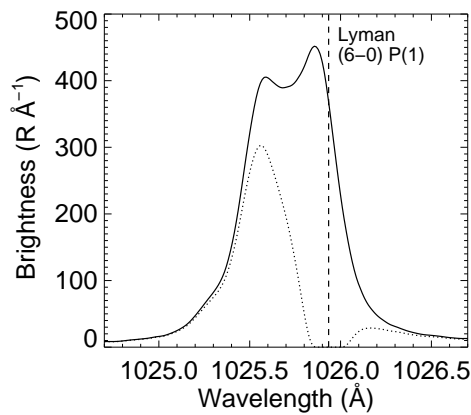}
}
\caption[SOHO/SUMER solar Lyman-$\beta$ profiles.]{SOHO/SUMER solar Lyman-$\beta$ profiles \citep[from ][]{Lemaire:2002} showing positions of the \Htwo\ Lyman (6$-$0)P(1) line and the absorption profile for a column density of 10$^{20}$~cm$^{-2}$. \label{sumer_lyb}}
\end{center}
\end{figure}
\subsection{$FUSE$ Spectra of Jupiter and Saturn}

A section of the Jovian spectrum, dominated by electron-impact excited bands of \Htwo, is shown in Figure~\ref{fusejup}. The electron impact spectrum was modeled using the method described in \citet{Wolven:1998}, with a photoelectron energy of 30~eV. The spectrum is well described by a model with an \htwo\ temperature of 700~K, and an \Htwo\ column of $1 \times 10^{20}$~\cmsq, as shown by the red line in Figure~\ref{fusejup}. The \htwo\ column density roughly corresponds to the depth in the atmosphere where CH$_{4}$ absorption becomes optically thick. The excess emission of the Q(3) line of the Werner (1$-$4) band at 1163.8~\AA\ over the electron excitation model is due to pumping by two solar lines, \ion{N}{3} $\lambda$989.79 and \ion{O}{6} $\lambda$1031.91. The Saturn spectrum, shown in Figure~\ref{specsat}, is dominated by the auroral emission, with the corresponding electron impact model for \trot~=~400~K overplotted in red. 

In both spectra, the most striking feature unaccounted for by electron impact excitation is the P(1) line of the Lyman (6$-$3) band at 1166.8 \AA, whose position is indicated in the figures. This line is broader, asymmetric, and its peak is shifted blueward relative to the electron excited features. The Ly(6$-$1)P(1) line at 1071.6~\AA\ and the Ly(6$-$2)P(1) line at 1118.6~\AA\ have a similar appearance. These lines are part of the fluorescent progression pumped by the solar Lyman-$\beta$ line overlapping the Ly(6$-$0)P(1) line at 1025.93~\AA. Their shape reflects the fact that there are more solar Lyman-$\beta$ photons available on the blue side of the Ly(6$-$0)P(1) line center than on its red side, as can be seen in Figure~\ref{sumer_lyb}. This reciprocity in the line profile can only be achieved if we allow the photons in the line wings to scatter coherently, preserving their energy offset from the line core. The coherent scattering process requires dropping the CRD assumption in the fluorescence model, and using the PRD matrix to compute the emerging line profile. We construct a fluorescence model for the \htwo\ molecule, and show that the use of PRD gives a good agreement with the data, while the CRD approximation is inconsistent with the observations. Since the electron-impact spectrum has been modeled separately, in the following section we will describe in detail the modeling of the fluorescent cascade following pumping by solar Lyman-$\beta$.

\section{FLUORESCENCE MODEL WITH PARTIAL FREQUENCY REDISTRIBUTION}{\label{secmodel}}

\subsection{Radiative Transfer}{\label{ratran}}

The $specific$ $intensity$ of radiation, I$_{\nu}$, is defined as the energy per unit frequency emitted per unit area in unit time, and in unit solid angle along the specified direction, in units of ergs~s$^{-1}$~cm$^{-2}$~sr$^{-1}$~Hz$^{-1}$. The emergent intensity at the surface of the medium, measured by the observer, is also called $surface$ $brightness$. The behavior of the specific intensity at any point in the medium is described by the radiative transfer equation
\begin{equation}
\textbf{n}\cdot\triangledown I_{\nu}=-\kappa_{\nu}I_{\nu}+j_{\nu},
\end{equation}
\noindent where $\kappa_{\nu}$ is the absorption coefficient (cm$^{-1}$), and j$_{\nu}$ is the emissivity (ergs s$^{-1}$ cm$^{-3}$ sr$^{-1}$~Hz$^{-1}$). This equation becomes
\begin{equation}
\mu\frac{dI_{\nu}}{dz}=-\kappa_{\nu}I_{\nu}+j_{\nu},
\end{equation}
for a plane-parallel atmosphere, $\mu$ being the cosine of the angle between the direction of the ray and the normal to the atmosphere. 
For every absorber, we can write the absorption coefficient in terms of the scattering cross-section at frequency $\nu$, as $\kappa_{\nu}=n\sigma_{\nu}$, where $n$ is the number density (cm$^{-3}$) of the population with cross-section $\sigma_{\nu}$. This term can represent either line or continuum absorption. In the case of line absorption, $n$ will be replaced by the number density of the absorber in the lower state of the transition ($n_{i}$). In general, $\kappa_{\nu}$ will be the sum of multiple such terms, corresponding to all the species and transitions taken into account. Defining the optical depth as
\begin{equation}\label{tau}
\tau_{\nu}(z)=\int_{0}^{z}\kappa_{\nu}(z')dz'=N\sigma_{\nu},
\end{equation}
where N is the column density (cm$^{-2}$) of the population with cross-section $\sigma_{\nu}$, and the cross-section is depth-independent, the radiative transfer equation becomes
\begin{equation}\label{rt}
\mu\frac{dI_{\nu}}{d\tau_{\nu}}=-I_{\nu}+S_{\nu},
\end{equation}
where $S_{\nu}=j_{\nu}/\kappa_{\nu}$ is the source function. To simplify the remaining equations, the following derivations are performed for $\mu=1$. The results can be easily generalized for an arbitrary incidence angle.

The absorption and emission coefficients are the sum of contributions from continuum scattering and individual lines. In the case of spectral lines, these coefficients are related to the transition probabilities by
\begin{equation}\label{coeff}
\begin{split}
\kappa_{ij}(\nu)&=(h\nu/4\pi)(n_{i}B_{ij}\phi(\nu)-n_{j}B_{ji}\psi(\nu))\\
j_{ij}(\nu)&=(h\nu/4\pi)n_{j}A_{ji}\psi(\nu)
\end{split}
\end{equation}
where $j$ denotes the upper level, $i$ the lower level, and the angle dependence has been omitted. The quantities $n_{j}$ and $n_{i}$ are the number densities of atoms/molecules in the upper and lower level, respectively, and are related through the statistical equilibrium equations. The transition probabilities are given by the Einstein coefficients $A_{ji}$, while the Einstein $B$ coefficients for absorption and stimulated emission are given by
\begin{equation}
B_{ji}=\frac{g_{j}}{g_{i}}B_{ij}=\frac{c^{2}}{2h\nu^{3}}A_{ji},
\end{equation}
with $g$'s being the statistical weights of the corresponding levels. 

The radiative transitions are not infinitely narrow at the transition wavelength, but are broadened by the finite lifetimes of the excited states, and by bulk molecular motions in the medium. The frequency dependence of the absorption and emission line profiles are given by $\phi(\nu)$ and $\psi(\nu)$, respectively.

The lifetimes of molecular levels, defined as the inverse of $A_{v'J'}=\frac{1}{\Gamma}=\sum_{v''} A_{v'v''}$, determine the natural width of the lines, or the Lorentz profile. The gaussian distribution of the molecular motions results in the Doppler profile, with the Doppler width given by $\Delta\nu_{D}=\nu_{0}$v$/c$, where v is the sum of thermal and turbulent motions, v$=\sqrt{\mathrm{v}_{T}^{2}+b^{2}}$. The thermal velocity is v$_{T}=\sqrt{2kT/m}$, while $b$ represents the microturbulent velocity, and will be referred as the $Doppler$ $parameter$. The combination of the natural and Doppler broadening results in a profile for the absorption line $\phi$ of the form
\begin{equation}\label{eq4}
\phi(\nu)=\frac{1}{\Delta\nu_{D}\sqrt{\pi}}H(a,u),
\end{equation}
where $H(a,u)$ is the Voigt function, with $a=\frac{\Gamma}{4\pi\Delta\nu_{D}}$ and u$=\frac{\nu-\nu_{0}}{\Delta\nu_{D}}$. 

In general, the emission profile $\psi$ is different from the absorption profile, $\phi$, for the same transition. The two profiles are interrelated through the redistribution function $R(\nu$,\textbf{n},$\nu'$,\textbf{n$'$}), that represents the probability distribution that the photon is absorbed at frequency $\nu$ in direction \textbf{n}, and re-emitted at $\nu'$ in direction \textbf{n$'$}. It satisfies the normalization condition
\begin{equation}\label{eq5}
\frac{1}{(4\pi)^{2}}\int\int\int\int R(\nu,\textbf{n},\nu',\textbf{n}')d\nu d\Omega d\nu' d\Omega'=1,
\end{equation}
and the emission and absorption profiles can be obtained as
\begin{equation}\label{eq6}
\begin{split}
\phi(\nu,\textbf{n})&=\frac{1}{4\pi}\int\int R(\nu,\textbf{n},\nu',\textbf{n}')d\nu' d\Omega',\\
\psi(\nu',\textbf{n}')&=\frac{1}{4\pi}\int\int R(\nu,\textbf{n},\nu',\textbf{n}')d\nu d\Omega.
\end{split}
\end{equation}
The redistribution functions have been introduced by \citet{Hummer1962} for resonance lines, and have been given later a broader foundation and extended to subordinate transitions by \citet{Milkey1975,Heinzel1981}, and \citet{Hubeny1982}. The redistribution function describes the coherent scattering of radiation in the rest frame of the emitter (atom or molecule), combined with the frequency incoherence determined by the motion of the emitter, the broadening of the energy levels, and the collisions with other particles. In this general scenario, the radiation is only partially redistributed (PRD) in frequency. The case when this process involves scattering in lines with the same upper level but different lower levels (subordinate), is referred to as cross-redistribution (XRD). 

A special case occurs when the absorption and emission frequencies are considered independent, equivalent to the complete frequency redistribution (CRD). With this assumption, the redistribution function takes the simple form
\begin{equation}\label{eq7}
R(\nu,\textbf{n},\nu',\textbf{n}')=\phi(\nu,\textbf{n})\phi(\nu',\textbf{n}'),
\end{equation}
and the emission profile is identical to the absorption profile. This approximation is the method of choice for small optical depths when the Doppler core of the line is predominant, and for forbidden lines occurring between sharp energy levels. However, this approximation has important consequences for the photon escape probabilities, and does not correctly predict the line profiles. The implications of using PRD $vs.$ CRD in the fluorescence models will be further explored in Section~\ref{genprd}.

Having defined the frequency redistribution functions, the general expression for the source function at frequency $\nu$, with multiple overlapping lines and continuum, will be \citep{Uiten1989,HubenyLites:1995}
\begin{equation}\label{sfunc}
S(\nu)=\frac{j_{c}(\nu)+\displaystyle\sum_{i,j}{(h\nu/4\pi)A_{ji}\phi_{ij}(\nu)\left[n_{j}+\displaystyle \sum_{k<j}{n_{k}Q_{kji}(\nu)}\right]}}{\kappa_{c}(\nu)+\displaystyle\sum_{i,j}{(h\nu/4\pi)B_{ij}\phi_{ij}(\nu)\left[n_{i}-\frac{g_{i}}{g_{j}}\left(n_{j}+\displaystyle \sum_{k<j}{n_{k}Q_{kji}(\nu)}\right)\right]}},
\end{equation}
where $Q_{kji}(\nu)$ is defined as
\begin{equation}
Q_{kji}(\nu)=\frac{B_{kj}}{P_{j}}\int\left[\frac{R_{kji}(\nu,\nu')}{\phi_{ij}(\nu)}-\phi_{kj}(\nu')\right]I(\nu')d\nu',
\end{equation}
with $P_{j}$ denoting the total probability per unit time for transitions out of level $j$ (radiative and collisional), and all the quantities being angle-integrated. The general form of the angle-averaged redistribution function $R_{kji}(\nu,\nu')$, taking into account collisional broadening, is
\begin{equation}\label{req4}
R_{kji}(\nu,\nu')=(1-\gamma)\phi_{kj}(\nu')\phi_{ij}(\nu)+\gamma R_{kji}^{II}(\nu,\nu'),
\end{equation}
where $\gamma$ is the coherence fraction, defined as $\gamma=P_{j}/(P_{j}+Q_{j})$, $Q_{j}$ being the rate of $elastic$ collisions for level $j$, and $P_{j}$ is the total depopulation rate of level $j$. 

We can show that for the \htwo\ transitions pumped by solar Ly$\beta$ in planetary atmospheres, the depopulation rate of level $j$ via allowed transitions is much higher than both forbidden and collisional rates. The observed \htwo\ emission is mainly coming from the upper parts of the atmosphere, above the altitude at which the methane optical depth becomes one. This corresponds to the regions of lowest density and highest temperature in the atmosphere. Computing a weighted average of the atmospheric models for Jupiter and Saturn, down to the methane optical depth limit, we obtain \htwo\ densities of 7$\times$10$^{12}$~cm$^{-3}$ and 10$^{12}$~cm$^{-3}$, and temperatures of 717~K and 350~K for Jupiter and Saturn, respectively. Using kinetic theory, the number of collisions per second between \htwo\ molecules is given by $n_{coll}=\sqrt{2}\pi d^{2} n \sqrt{8kT/\pi m}$, where $d$ is the diameter of \htwo\ molecules (2.4~\AA), $n$ is the number density, $k$ is the Boltzmann constant, $T$ is the temperature, and $m$ is the mass of the \htwo\ molecules. This expression gives us an estimate of 5$\times$10$^{3}$~collisions/s for Jupiter and 5$\times$10$^{2}$~collisions/s for Saturn, including both elastic and inelastic collisions, between all the hydrogen molecules, regardless of their energy or excitation state. This estimate can be considered an upper limit for both the elastic collision rate $Q_{j}$ and the inelastic collision rate $C_{j}$ affecting level $j$. We can write $P_{j}= R_{j}+C_{j} =A_{j}+F_{j}+C_{j}$, where $R_{j}$ is the sum of the radiative rates, $C_{j}$ the sum of the collisional rates ($<$10$^{3}$~s$^{-1}$), and $A_{j}$ the sum of the allowed radiative rates ($\sim$10$^{9}$~s$^{-1}$), and $F_{j}$ the sum of the forbidden radiative rates ($\sim$10$^{-6}$~s$^{-1}$). Since $Q_{j}$ is also of the order 10$^{3}$~s$^{-1}$, with good approximation $P_{j}\simeq A_{j}\gg Q_{j}$. Following this argument, in the fluorescence model presented in Section~\ref{fluormod} we assume $\gamma=1$. However, for comparison purposes, we also compute the model in the CRD approximation, corresponding to $\gamma=0$. The results for the two limiting cases are shown in Section~\ref{data}.

\begin{figure}
\begin{center}
\epsscale{1.05}
\rotatebox{0}{
\plotone{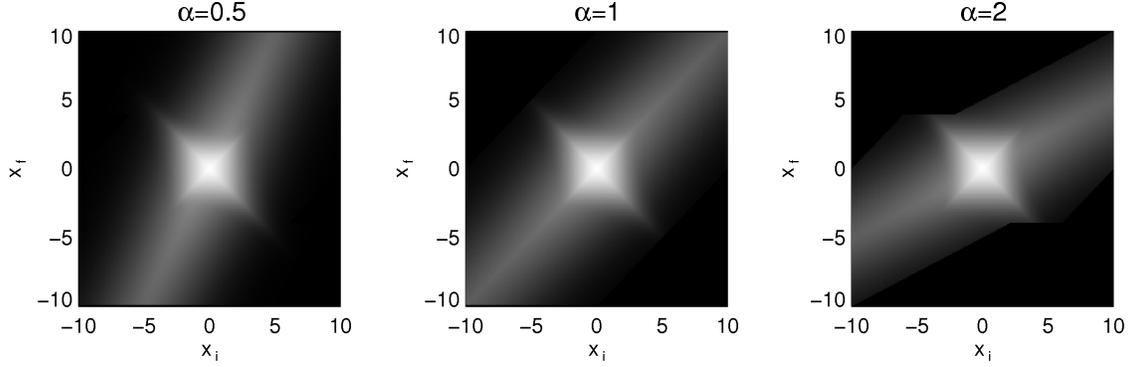}
}
\caption[Frequency redistribution matrices.]{ Frequency redistribution matrices computed for three values of $\alpha$, as labeled. The redistribution function is constructed for $v_{D}$=1~km~s$^{-1}$ and $a=0.01$, at a resolution of 0.1 Doppler widths.\label{matrix}}
\end{center}
\end{figure}

In the last term in Eq.~\ref{req4}, $R_{kji}^{II}(\nu,\nu')$ is the angle-averaged redistribution function derived by \citet{Milkey1975,Hubeny1982} for transitions with a common upper level and sharp lower levels:
\begin{equation}\label{r2}
R^{II}(x_{1},x_{2})=\pi^{-3/2}\int_{u^{-}}^{\infty}e^{-u^2}\left[arctan\left(\frac{\bar{y}}{a_{1}}\right)-arctan\left(\frac{\underline{y}}{a_{1}}\right)\right]du,
\end{equation}
where $x_{1}$ and $x_{2}$ are the distances from the line centers of the absorbing and emitting lines, respectively, in units of Doppler widths (~$x\equiv(\nu-\nu_{0})/\Delta\nu_{D}$~), $a_{1}$ is the damping parameter of the absorbing line (~$a\equiv\Gamma/4\pi\Delta\nu_{D}$, $\Gamma=A_{j}+2Q_{j}$~), and the other quantities are defined as
\begin{equation}
\begin{aligned}
\underline{y}&\equiv max(x_{1}-u,\alpha(x_{2}-u)),& u^{-}&\equiv\frac{|x_{1}-\alpha x_{2}|}{1+\alpha},\\
\bar{y}&\equiv min(x_{1}+u,\alpha(x_{2}+u)),& \alpha&\equiv\frac{\lambda_{1}}{\lambda_{2}}.\\
\end{aligned}
\end{equation}
The redistribution matrices depend on the ratio of the central wavelengths of the absorbing and emitting lines, $\alpha$, and on the temperature and turbulent velocities through the Doppler broadening. Therefore, new matrices have to be computed for each pair of transitions, and at each depth in the slab, if the Doppler broadening is variable. The number of matrix elements to be evaluated is reduced in half by the symmetry property $R^{II}(x_{1},x_{2})=R^{II}(-x_{1},-x_{2})$. The form of the redistribution matrices for three values of $\alpha$ is shown in Figure~\ref{matrix}. The color scale stretch in the figure has been chosen to emphasize the behavior in the line wings. We distinguish two regimes, one in the line core, where the photons are redistributed uniformly forming the Gaussian profile, and another in the line wings, where the photons are coherently scattered, preserving the energy difference from the line center. The CRD case can be visualized as the central bright square region. It is clear that expanding this behavior beyond the central $\pm$2 Doppler units will result in large errors.

The radiative transfer equation can be solved formally for a medium of total optical depth $T_{\nu}$, yielding
\begin{equation}\label{formsol}
\begin{split}
I_{\nu}^{+}(\tau_{\nu})&=I_{\nu}^{+}(T_{\nu})e^{-(T_{\nu}-\tau_{\nu})}+\int_{\tau_{\nu}}^{T_{\nu}}e^{-(t-\tau_{\nu})}S_{\nu}(t)dt,\\
I_{\nu}^{-}(\tau_{\nu})&=I_{\nu}^{-}(0)e^{-\tau_{\nu}}+\int_{0}^{\tau_{\nu}}e^{-(\tau_{\nu}-t)}S_{\nu}(t)dt,
\end{split}
\end{equation}
for the reflected and transmitted parts of the intensity, respectively. This expressions can be evaluated directly for each frequency when the source function at all the points in the medium is known, using either the differential Feautrier method \citep{Feautrier1964,Mihalas:1978}, or the integral short characteristics method of \citet{Kunasz1988}. However, in most cases the source function is itself dependent on the local specific intensity, both through the level populations and the emission profiles, as seen from Eq.~\ref{sfunc}. 

Even for the cases where the radiative transfer equation can be linearized, the resulting system contains all the depth-dependent frequency points and their cross-correlations through the spatial derivatives and the redistribution matrices. The direct inversion of such a system is numerically prohibitive. As a consequence, the method of solution proceeds iteratively, starting with an initial estimate of the source function to determine the specific intensities and level populations, which are in turn used to obtain the new estimate for the source function. This method, known as $\Lambda$-iteration, has less memory requirements, but the computing time per iteration is dependent on the number of depth and frequency points, and its convergence is slow for large optical depths. The alternative solution used in this work is based on the approximation by \citet{LiuDalgarno:1996}, that does not require multiple iterations or matrix inversions, but relies on a number of simplifying assumptions. This method is presented in detail in the next section.

\subsection{Fluorescence Model}{\label{fluormod}}

The starting model for the treatment of \htwo\ fluorescence, described in Section~\ref{genform}, was introduced by \citet{LiuDalgarno:1996}, and is based on a set of simplifying assumptions that will be used throughout the paper. The model equations are derived for a plane-parallel atmosphere, neglecting the stimulated emission term in the absorption coefficient. The distribution of molecules in the ground state is assumed to be a Maxwell-Boltzmann distribution given by
\begin{equation}\label{eeq9}
N_{vJ}=N_{tot}\frac{(2J+1)g_{s}exp(-E_{vJ}/kT)}{\sum_{v,J} (2J+1)g_{s}exp(-E_{vJ}/kT)},
\end{equation}
where $N_{tot}$ is the total column density, $E_{vJ}$ is the energy in the ($v$\arcsec,~$J$\arcsec) level, T is the temperature, and $g_{s}$ is the nuclear spin statistical weight. The weight $g_{s}$ is 3 for odd rotational quantum numbers of H$_{2}$, and 1 otherwise. Unless otherwise specified, the quoted temperature is used for deriving both the Doppler line width and the ground state distribution. This distribution can be easily replaceable with a user-specified one for special cases. The medium is characterized by global temperatures and column densities for all the constituent species.

The energy difference between the electronic ground state and the excited electronic states is large enough that the population of the upper levels is determined solely by the radiative excitation, except when electron impact excitation is present. The upper levels of the  \htwo\ transitions pumped by solar Ly$\beta$ belong to the excited electronic state B$^{1}\Sigma_{u}^{+}$, 11~eV above the ground electronic state. At the temperatures in the upper atmosphere of Jupiter and Saturn (300-700~K), the average energy of the \htwo\ molecules is 0.03-0.06 eV, insufficient to excite the molecules into the B$^{1}\Sigma_{u}^{+}$ state. Even though in most parts of the atmosphere the dominant collisions are with molecular hydrogen, these collisions will not affect the population of the excited electronic states, because the hydrogen molecules don't have enough energy to excite these states, and the collisional frequency between \htwo\ molecules is much smaller than the lifetimes of the excited electronic states. Due to the short lifetimes of the B$^{1}\Sigma_{u}^{+}$ state levels ($\sim$10$^{-9}$ seconds), it is unlikely that collisions with electrons or hydrogen molecules (of the order of 10$^{3}$ per second, or less) will change the population of these levels before they de-excite into the X$^{1}\Sigma_{g}^{+}$ state via a radiative transition. Excitations and de-excitations between neighboring levels in the B$^{1}\Sigma_{u}^{+}$ state are dipole-forbidden, and therefore they have a negligible contribution to the level populations. The fluorescence rates are also assumed small, so that the distribution in the ground electronic state X$^{1}\Sigma_{g}^{+}$ satisfies thermal equilibrium, eliminating the need for solving the detailed balance equations. This assumption is adequate for planetary atmospheres, but breaks down in the ISM regions where far-UV fluorescence is the dominant factor determining the ground state population of \htwo. Adopting this approximation in general, we can predict the effects of PRD on the emergent line shapes, but only approximately the relative line strengths.

While the photoelectrons (tens of eV, on average) and the auroral electrons (in the 10-100 keV range) have enough energy to excite transitions into the B$^{1}\Sigma_{u}^{+}$ state, we have taken this contribution into account separately, as it's responsible for most of the far-UV \htwo\ emission observed in planetary atmospheres \citep{Shemansky:1983,LiuDalgarno:1996}. The electron-impact excitation model is not restricted to a handful of lines (like the Ly$\beta$ pumping model), but includes all measured \htwo\ transitions in the Lyman (B$^{1}\Sigma_{u}^{+}$-X$^{1}\Sigma_{g}^{+}$), Werner (C$^{1}\Pi_{u}$-X$^{1}\Sigma_{g}^{+}$), Lyman cascade (EF$^{1}\Sigma_{g}^{+}$-B$^{1}\Sigma_{u}^{+}$-X$^{1}\Sigma_{g}^{+}$), as well as the B$'$$^{1}\Sigma_{u}^{+}$-X$^{1}\Sigma_{g}^{+}$, D$^{1}\Pi_{u}$-X$^{1}\Sigma_{g}^{+}$, B\arcsec$^{1}\Sigma_{u}^{+}$-X$^{1}\Sigma_{g}^{+}$, and D$'^{1}\Pi_{u}$-X$^{1}\Sigma_{g}^{+}$ systems. This electron-impact excitation model is shown in Figures~\ref{fusejup} and \ref{specsat} for Jupiter and Saturn, respectively. A similar model for continuum-pumped fluorescence, including all far-UV \htwo\ electronic transitions, is especially prohibitive, and we include a brief exemplification in Section~\ref{h2nebmod}.

\subsubsection{Single Scattering Approximation}{\label{genform}}

As can be seen from Eq.~\ref{formsol}, in the absence of emitters in the medium, the emergent intensity through a slab of optical depth $\tau_{\nu}$ is $I_{\nu}(\tau_{\nu})=I_{\nu}(0)e^{-\tau_{\nu}}$. A medium is considered optically thick at frequency $\nu$ when $\tau(\nu)\geqslant 1$, and the transmitted radiation intensity is attenuated by a factor of $e$. Let ${\cal G}_{0}(\nu)$ denote the specific intensity of the incident solar radiation field. The total amount of radiation absorbed by the slab will be given by
\begin{equation}\label{absflx}
{\cal G}(\nu)={\cal G}_{0}(\nu)\left(1-e^{-\tau(\nu)}\right)
\end{equation}
\noindent where $\tau(\nu)$ is the $total$ optical depth summed over all possible transitions and continuum,
\begin{equation}\label{eeq7}
\tau(\nu)=\tau_{c}(\nu)+\sum_{i,j} \tau_{ij}(\nu).
\end{equation}
In this manner, the method accounts for line saturation and line overlaps, and multiple species can contribute to $\tau(\nu)$. The frequency dependence of the line optical depth is characterized by the Voigt profile. In the absence of pure absorption, photon number conservation requires that the quantity in Eq.~\ref{absflx}, integrated over all frequencies, be the same as the total integrated intensity of the fluorescent lines. The fraction of the absorbed intensity from Eq.~\ref{absflx} that is due to a particular transition ($k\rightarrow j$) is $\tau_{kj}(\nu)/\tau(\nu)$. To obtain the rate at which the upper level $j$ is populated through the transition ($k\rightarrow j$), we multiply this fraction with ${\cal G}(\nu)$ and integrate over all frequencies. Denoting $I^{obs}$ the $emergent$ intensity at the top of the atmosphere, the total intensity in the fluorescent line ($j\rightarrow i$) will be given by
\begin{equation}\label{eeq11}
I_{ji}^{obs}=\left(1-\eta_{j}\right)\displaystyle\sum_{k}\left(\int {\cal G}(\nu)\frac{\tau_{kj}(\nu)}{\tau(\nu)}d\nu\right)\frac{A_{ji}}{A_{j}}
\end{equation}
where $\eta_{j}$ is the pre-dissociation fraction of the upper level, and the last fraction is the branching ratio for the transition in line ($j\rightarrow i$) to occur. In molecules, the excited electronic states can overlap with the dissociation continuum of another electronic state, leading to radiationless transitions between these two states, process called pre-dissociation \citep{Herzberg1950}. This process affects strongly some of the excited electronic states of the hydrogen molecule, such as B\arcsec $^{1}\Sigma_{u}^{+}$, D $^{1}\Pi_{u}$, D\arcmin $^{1}\Pi_{u}$, and D\arcsec~$^{1}\Pi_{u}$ \citep{Ajello:1984}. A fraction of the population of these states will therefore be lost without a radiative de-excitation. This loss is accounted for by the pre-dissociation fraction term in Eq.~\ref{eeq11}. A more in-depth justification of Eq.~\ref{eeq11} and its relation to the solution of the radiative transfer equation will be presented in Section~\ref{genprd}.

This basic treatment of the emission process does not take into account multiple scattering in optically thick lines, and cannot reproduce the effects of self-absorption. A method to reproduce the self-absorption effects for optically thick CO lines has been described in \citet{Lupu:2007}. Due to the integration over frequencies in Eq.~\ref{eeq11}, all the information about the line profile is lost, and the result is equivalent to using the CRD approximation in the source function, given that the Doppler core of the lines is not resolved. In the next section we extend this framework to allow for depth variations of the model parameters, and to model the emission line profiles using partial frequency redistribution.

\subsubsection{Treatment of Frequency Redistribution}{\label{genprd}}

The PRD radiative transfer problem is more complex than the CRD case, due to the heavy couplings in frequency in addition to spatial correlations and detailed balance \citep{HubenyLites:1995,Uiten:2001}. Even restricting the geometry to the plane-parallel case, and assuming fixed level populations, the problem remains computationally expensive due to intrinsically large optical depths and fine frequency grids. 

In this section we develop a multilayer extension of the method of \citet{LiuDalgarno:1996} presented above, that reproduces the effects of multiple scattering and the emission line profiles using the frequency redistribution matrix. The basic idea is that one uses an equation of the same form as Eq.~\ref{eeq11} to calculate the specific intensities at each depth in the medium, and then propagate the outgoing intensity to the surface by adding these contribution to the next layer, step-by-step.

We will work in the approximation that stimulated emission is negligible, and that the number density in the upper level $n_{j}$ is determined only by the radiative transitions (see also Eq.~\ref{nreq}). These approximations are justified by the short lifetime of the upper level $j$ of the transitions pumped by solar Ly$\beta$. The levels higher than $j$ do not contribute significantly to the population of level $j$ via spontaneous emission, since transitions between different ro-vibrational levels of the B$^{1}\Sigma_{u}^{+}$ state are forbidden in the dipole approximation, while the electronic transitions between the B$^{1}\Sigma_{u}^{+}$ and X$^{1}\Sigma_{g}^{+}$ states are dipole-allowed. The difference between the transition probabilities of the dipole-forbidden and the dipole-allowed transitions is about 14 orders of magnitude, and therefore a spontaneous emission out of a ro-vibrational level in the B$^{1}\Sigma_{u}^{+}$ state is much more likely to go to the ground electronic state X$^{1}\Sigma_{g}^{+}$ than to another level within the B$^{1}\Sigma_{u}^{+}$ state. However, the population of the B$^{1}\Sigma_{u}^{+}$ state can also be modified via cascade from the EF$^{1}\Sigma_{g}^{+}$ states. The EF$^{1}\Sigma_{g}^{+}$ states are populated via electron impact excitation from the ground electronic state, as the transitions between these two levels (X and E,F) is also dipole forbidden. This cascade process has been taken into account in the electron impact excitation model, computed separately, and added to the fluorescence model for the final spectrum.

Although the energy difference between the ro-vibrational levels in the B$^{1}\Sigma_{u}^{+}$ state is small enough, such that collisional transitions between such levels could be possible given the energy of the collision partners in planetary atmospheres, the lifetime of such ro-vibrational levels in the B$^{1}\Sigma_{u}^{+}$ state ($\sim 10^{-9}$~s) is too short for collisional excitation and de-excitation to occur. As discussed in Section~\ref{ratran}, under the parameters describing planetary atmospheres, the upper limit for the average collision rate beween \htwo\ molecules is about 10$^{2-3}$~s$^{-1}$, much smaller than the radiative de-excitation rate for a dipole-transition, which is about 10$^{7}$~s$^{-1}$. Since this estimate for the collision rate does not take into account the kinetic energy of the partners, nor the excitation state of the \htwo\ molecules, the actual collisional transition rate between any two excited levels in the B$^{1}\Sigma_{u}^{+}$ state will be much smaller. 

Under these approximations, we can write for the population of level $j$ determined by radiative excitation
\begin{equation}\label{req1}
\begin{split}
n_{j}&=\displaystyle\sum_{k<j}\frac{n_{k}B_{kj}}{P_{j}}\int\phi_{kj}(\nu')I(\nu')d\nu',\\
\kappa(\nu)&=\kappa_{c}(\nu)+(h\nu/4\pi)B_{ij}\phi_{ij}(\nu)n_{i}.
\end{split}
\end{equation}
After cancellations, the source function will be
\begin{equation}\label{req2}
S(\nu)=\displaystyle\sum_{i,j}\frac{h\nu}{4\pi}\frac{A_{ji}}{\kappa(\nu)}\displaystyle\sum_{k<j}\frac{n_{k}B_{kj}}{P_{j}}\int R_{kji}^{II}(\nu,\nu')I(\nu')d\nu',
\end{equation}
where $R_{kji}^{II}$ is the general redistribution function (Eq.~\ref{r2}) for coherent scattering in the atom's frame and negligible lower level broadening (metastable state) \citep{Hubeny1982,Hummer1962}. A generalization of this expression, including the effects of collisions (negligible in our case), is presented in the Appendix.

The radiative transfer Eq.~\ref{rt}, with the source function given by Eq.~\ref{req2}, can now be solved  using one of the methods mentioned in Section~\ref{ratran}. We investigated the feasibility of these methods for the full treatment of the H$_{2}$ molecule, but the large number of transitions (of the order 10$^{4}$) and the fine frequency and depth grids, renders them impractical due to the large memory requirements and large number of iterations. On the other hand, the approximation of \citet{LiuDalgarno:1996} described in Section~\ref{genform} is very fast, but suffers from the inability to reproduce the line profiles and self-absorption. We are able to show that the PRD process can be easily introduced in this formalism, and that the effects of multiple scattering are reproduced by dividing the medium in multiple layers, and applying this method layer-by-layer.

We divide the medium into $M$ layers of constant parameters (number densities, temperatures, etc.), indexed by $l$, starting with 0 at the surface. The formal solution for the inward radiation (Eq. \ref{formsol}) for one layer then becomes
\begin{equation}\label{di}
I^{-}(\tau^{l+1},\nu)=I^{-}(\tau^{l},\nu)e^{-\Delta\tau(\nu)}+e^{-\Delta\tau(\nu)}\int_{\tau^{l}}^{\tau^{l+1}}e^{t}S(t)dt\equiv I^{-}(\tau^{l},\nu)e^{-\Delta\tau(\nu)}+I_{d}(\tau^{l+1},\nu),
\end{equation}
where $\Delta\tau=\tau^{l+1}-\tau^{l}$ and I$_{d}(\tau^{l+1})$ represents the local diffuse radiation field. Denoting the incident intensity by ${\cal G}_{0}(\nu')$, each layer will be illuminated by ${\cal G}_{0}(\nu')e^{-t(\nu')}$. Substituting this initial value in the the source function from Eq.~\ref{req2}, we obtain
\begin{equation}
I_{d}(\tau^{l+1},\nu)=\displaystyle\sum_{i,j}\frac{A_{ji}}{P_{j}}\displaystyle\sum_{k<j}\frac{h\nu}{4\pi}n_{k}B_{kj}e^{-\Delta\tau}\int_{\tau^{l}}^{\tau^{l+1}}\frac{e^{t(\nu)}}{\kappa(\nu)}\int R_{kji}^{II}(\nu,\nu'){\cal G}_{0}(\nu')e^{-t(\nu')}d\nu'dt,
\end{equation}
or, collecting the optical depth dependent terms,
\begin{equation}
I_{d}(\tau^{l+1},\nu)=\displaystyle\sum_{i,j}\frac{A_{ji}}{P_{j}}\displaystyle\sum_{k<j}\frac{h\nu}{4\pi}n_{k}B_{kj}\int R_{kji}^{II}(\nu,\nu'){\cal G}_{0}(\nu')\left[e^{-\Delta\tau}\int_{\tau^{l}}^{\tau^{l+1}}\frac{e^{t(\nu)-t(\nu')}}{\kappa(\nu)}dt(\nu)\right]d\nu'.
\end{equation}

Making the usual discretization approximation where all parameters are constant for each layer, $n_{k}$ will represent the number density in level $k$ for the layer $l$, and $\kappa(\nu)$ the absorption coefficient for the same layer. We can also write, for short, $\kappa(\nu)=\kappa$, $\kappa(\nu')=\kappa'$, $t(\nu)=\kappa(\nu)z=\kappa z$, $t(\nu')=\kappa(\nu')z=\kappa' z$. The optical depth integral outlined in the square brackets then yields
\begin{equation}
K=\frac{e^{-\kappa(z^{l+1}-z^{l})}}{\kappa-\kappa'}\left[e^{z^{l+1}(\kappa-\kappa')}-e^{z^{l}(\kappa-\kappa')}\right]\xrightarrow[\kappa=0]{}\frac{e^{-\kappa'z^{l}}}{\kappa'}\left[1-e^{-\kappa'(z^{l+1}-z^{l})}\right].
\end{equation}
Separating the emission process from the absorption process, in the last expression we made the approximation that the absorption coefficient will be 0 at the emitted frequency $\nu$, such that the equation for the local intensity becomes linear:
\begin{equation}\label{di2}
I^{-}(\tau^{l+1},\nu)=I_{d}(\tau^{l+1},\nu)=\displaystyle\sum_{i,j}\frac{A_{ji}}{P_{j}}\displaystyle\sum_{k<j}\frac{h\nu}{4\pi}\Delta N_{k}B_{kj}\int R_{kji}^{II}(\nu,\nu')\frac{{\cal G}_{0}(\nu')e^{-\tau^{l}(\nu')}}{\Delta\tau(\nu')}\left[1-e^{-\Delta\tau(\nu')}\right]d\nu'.
\end{equation}
In the last step we introduced the column density in level $k$ for layer $l$, $\Delta N_{k}=n_{k}\Delta z=n_{k}(z^{l+1}-z^{l})$, and replaced the absorption coefficient by the optical depth of the layer ($\Delta\tau=\kappa\Delta z$) in the denominator.

In the absence of collisions, $P_{j}$ is the sum of all radiative transitions from level $j$, containing both the emission and absorption rates. In our case, we do not have absorption from the excited electronic states, so that $P_{j}$ reduces to the sum of all emission transition probabilities, or $P_{j}=A_{j}$. We also need to multiply this expression with the pre-dissociation factor ($1-\eta_{j}$). In the CRD approximation, when the redistribution function takes the form in Eq.~\ref{eq7}, the expression for intensity Eq.~\ref{di2} reduces to the \citet{LiuDalgarno:1996} form in Eq.~\ref{eeq11}, after taking into account Eqs.~\ref{coeff} and \ref{tau} for the line optical depth.

In our multilayer method, the Eq.~\ref{di2} is only the initial estimate of the intensities at each spatial point. In the next step, we start propagating these intensities outward, assuming total reflection at the bottom boundary. In this process we account for absorption and re-emission of the outgoing radiation in each layer, writing
\begin{equation}\label{di3}
\begin{split}
I^{+}(\tau^{l},\nu)=&I_{d}(\tau^{l},\nu)+I_{d}(\tau^{l+1},\nu)e^{-\Delta\tau}\\
&+\displaystyle\sum_{i,j}\frac{A_{ji}}{P_{j}}\displaystyle\sum_{k<j}\frac{h\nu}{4\pi}\Delta N_{k}B_{kj}\int R_{kji}^{II}(\nu,\nu')\frac{I_{d}(\tau^{l+1},\nu')}{\Delta\tau(\nu')}\left[1-e^{-\Delta\tau(\nu')}\right]d\nu'.
\end{split}
\end{equation}
This expression ensures photon number conservation and progressively adds the contribution of each layer to the total emergent intensity. The resulting spectrum is computed in a single iteration, and does not require matrix inversions. The computing time depends mostly on the number of grid points in optical depth and the number of transitions for which we have to compute the redistribution matrices. 

 \begin{deluxetable}{cccc}
\tabletypesize{\small}
\tablecaption{Characteristics of the radiative transfer methods compared. \label{rttable}}
\tablewidth{0pt}
\tablehead{
& \colhead{Lambda Iteration} & \colhead{Multilayer}   & \colhead{Single} \\ 
& (Feautrier solution) & & layer }
\startdata
Resources & High & Moderate & Low\\
Consistency & Can be unstable & Constrained & Constrained\\
Convergence & Scales with optical depth & 2-pass layer-by-layer & Single step\\
Spatial variations  & Yes & Yes & No \\
\enddata
\end{deluxetable}

 \begin{figure}[ht]
\begin{center}
\epsscale{0.8}
\rotatebox{0}{
\plotone{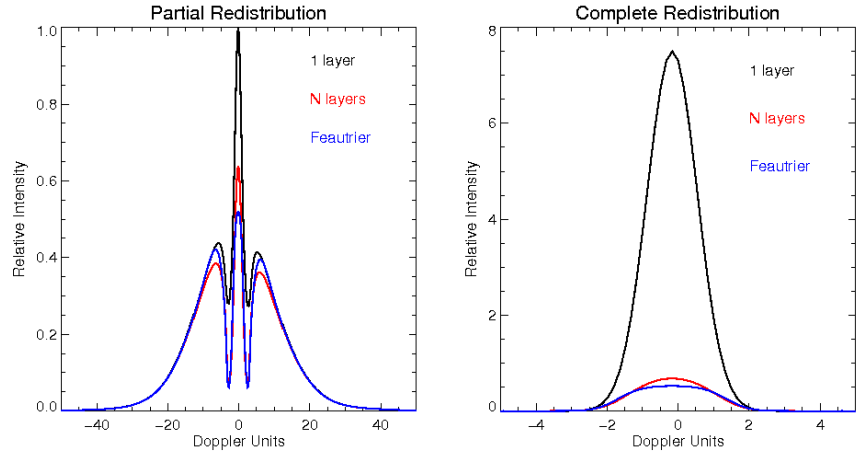}
}
\caption[Comparison of the emergent intensity under different computational approaches for the radiative transfer.]{ Comparison of the three methods of calculating the emergent intensity of Ly$\beta$ for an optical depth at line center of 4$\times$10$^{5}$, using the partial redistribution (PRD, left) and the complete redistribution (CRD, right) scenarios.\label{compprd}}
\end{center}
\end{figure}

The properties of this method versus the approximation of \citet{LiuDalgarno:1996} (single-layer), and the Feautrier solution are shown in Table~\ref{rttable}. The output from the three methods is illustrated in Figure~\ref{compprd}, for both the PRD and CRD cases. This comparison has been performed for a uniform medium of neutral hydrogen, with a total column density of 10$^{20}$~cm$^{-2}$ and a temperature of 700~K, corresponding to an optical depth at line center of 4$\times$10$^{5}$. The frequency grid was centered on Ly$\beta$, with a bin size of 5.44~GHz, and the optical depth grid was spaced using 10 points per dex at line center. Taking into account the instrumental resolving power, the results from the multilayer and Feautrier methods are effectively indistinguishable. The difference in the total integrated brightness of the line is about 2\% for the PRD case, and less than 10\% for CRD, while the computing times differ by a factor of about 100. The single layer approximation overpredicts by a large amount the number of photons in the line core, illustrating its inability to account for multiple scatterings (self-absorption). The decrease in the Ly$\beta$ intensity seen in the other cases is due to its branching ratio of 0.88.

\subsubsection{Effects of PRD Compared to CRD}{\label{scomp}}

As Figure~\ref{compprd} shows, the use of the PRD matrix results in a large effect on the line shape, due to a large number of photons escaping in the line wings. This effect becomes significant when the optical depth at line center exceeds 10$^{3}$, and is illustrated by the $FUSE$ observations of the atmospheres of Jupiter and Saturn presented in the next section. Large variations in the intensity of the exciting radiation over the line profile can also lead to a shift of the peak brightness from the line center, when the largest number of photons is absorbed in the line wing and re-emitted coherently.

 \begin{figure}[ht]
\begin{center}
\includegraphics[width=4.0in,angle=0,clip]{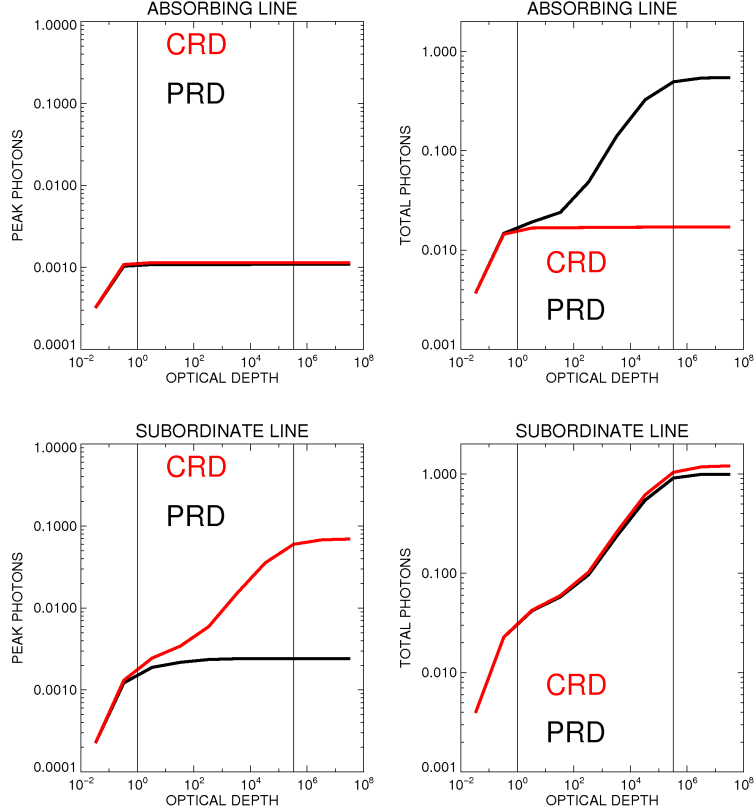}
\caption[The effects of PRD and CRD on the integrated line strengths and the peak number of photons.]{ The dependence of the number of photons in the central wavelength bin (left) and of the total number of photons integrated over the line profile (right) on the optical depth for an optically thick line (upper panels) and an optically thin line with the same upper level (lower panels). The results from the models using PRD and CRD are shown in black and red, respectively.\label{thickthin}}
\end{center}
\end{figure}

 \begin{figure}[h]
\begin{center}
\includegraphics[width=2in,angle=90,clip]{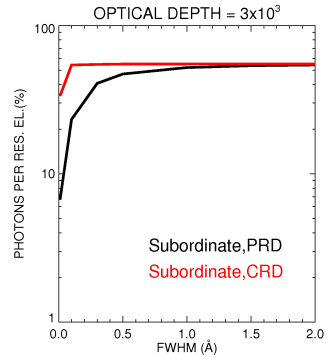}
\caption[The number of photons per resolution element at line center, in the PRD and CRD cases.]{The dependence of the measured signal (number of photons per resolution element) on the instrumental FWHM. The integration using the model with PRD is shown in black, and the one for the CRD case is shown in red. \label{resolve}}
\end{center}
\end{figure}

The use of PRD has even stronger consequences for the predicted integrated line strengths and the peak intensity at the line center, as illustrated by Figure~\ref{thickthin}. The panels show the dependence with total optical depth of the emergent number of photons in the central wavelength bin (left) and the integrated number of photons in the line (right), as predicted by the PRD and CRD scenarios (in black and red, respectively). The behavior of an optically thick line, where the absorption takes place for a set of transitions with a common upper level, is plotted in the upper set of panels, while the lower ones display the results for an optically thin line with the same upper level. To guide the eye, two vertical lines mark the points where the optical depth is one, and where the column density is 10$^{21}$~cm$^{-2}$, respectively. The transitions are pumped by the solar Ly$\beta$ profile normalized to an integrated intensity of 1. The saturation of the fluorescent line brightness at high optical depths in Figure~\ref{thickthin} reflects the absorption of all available photons from the incident radiation.

First, we see that the total brightness of the optically thick absorbing line is greatly increased by using PRD, due to the fact that more photons are able to escape in the line wings, without being re-emitted in an optically thin line from the same progression. In the CRD scenario, more photons are found at the line centre and undergo multiple scatterings, increasing the number lost to other transitions. It also shows that in the CRD case the integrated intensity remains roughly constant beyond unit optical depth, as expected.  
Correspondingly, the number of photons in the central wavelength bin is greatly suppressed for the optically thin subordinate lines. This fact is determined also by the coherent scattering in the line wings, as fewer photons will be redirected towards the line core. The total intensity is also somewhat decreased in the PRD case, by comparison with CRD, but this effect is not as noticeable on individual lines, because the scattered photons are shared among many transitions with the same upper level. The peak brightness at the line centre in the optically thick lines remains basically unaffected, reflecting the identity of the PRD and CRD matrices in the line core.

When the PRD treatment is employed, both subordinate (optically thin) and resonant (optically thick) lines show lower intensities at the line centre, and the line profiles are widened by the photons escaping in the line wings. This process results in a decrease of the line contrast across the spectrum, and is investigated in Section~\ref{h2nebmod}, in relation to the search for fluorescent emission from molecular hydrogen in reflection nebulae. Depending on the spectral resolution of the instrument, the PRD assumption predicts that the lines will be more difficult to detect than expected based on the estimates from CRD calculations. This difference, as shown by Figure~\ref{thickthin}, will significantly affect the subordinate lines. We estimate the effect on the line detection ability by integrating the number of photons from the optically thin subordinate line in a resolution element equal to the FWHM of the spectral resolution of the instrument. Figure~\ref{resolve} shows the resulting number of photons versus resolving power for an optical depth at the wavelength of the absorbing line of 3$\times$10$^{3}$, for both PRD (black) and CRD (red) models. The model predictions become indistinguishable when the instrumental FWHM is larger than 1\AA\ (R$<$1000), but we should expect a factor of 5 drop in the signal due to PRD when the instrumental FWHM is 0.01\AA\ (R=100,000).

The next section shows specific examples where the use of PRD is required to explain the observations, and the multilayer method is used to generate synthetic spectra that reproduce the line profiles for planetary atmospheres. As a generalization, we use a toy model to estimate the effects of PRD on the continuum-pumped \htwo\ fluorescence, with application to high resolution observations of reflection nebulae.
\section{RESULTS}{\label{data}}

\subsection{Planetary Atmospheres}

For both planets observed by $FUSE$ we compute the \htwo\ fluorescence spectrum pumped by solar Lyman-$\beta$ using the multilayer approach with depth dependent parameters. The variation of gas composition and temperature with altitude was determined from current atmospheric models. The species with contributing opacities were \ion{H}{1}, \Htwo, and CH$_{4}$. For the exciting radiation field we used the solar Lyman-$\beta$ flux from the compilation of \citet{Lemaire:2002}, measured on a date close to that of the \fuse\ observations.

\begin{figure}
   \begin{center}  
   \includegraphics[angle=0,width=5in,clip]{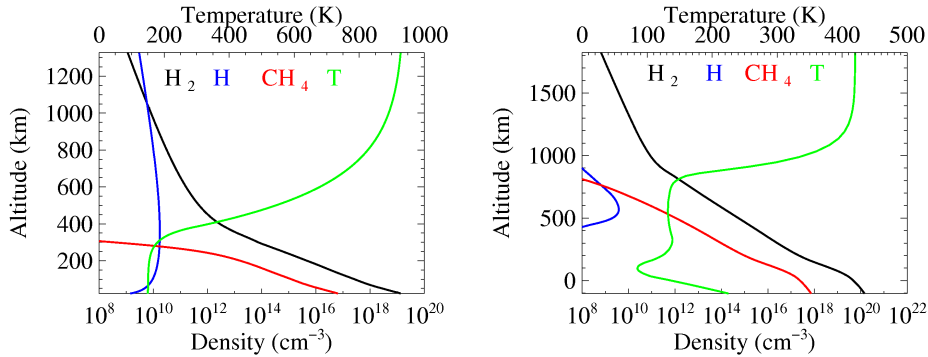}
   \end{center}
   \caption[Atmospheric models for Jupiter and Saturn. ]{\label{jupatm} 
The atmospheric models of Jupiter (left) and Saturn (right) used for the synthetic fluorescence spectrum. }
   \end{figure} 

\subsubsection{Jupiter}

The atmospheric model for Jupiter is shown in the left panel of Figure~\ref{jupatm}. The fluorescence model was integrated for a set of 19 lines of sight through the atmosphere, starting with the innermost region falling into the slit, at 13\farcs66 from the planet center, and continuing with a logarithmic spacing up to the outer radius. The depth variation of the model parameters was adjusted for each beam, accounting for the variation of layer thicknesses and composition with projected distance from the disk center. The observed integrated brightness was also corrected for the slit filling fraction during the exposure, as the slit was scanning the limb (see Figure~\ref{obsjup}). The effective area of the disk observed is given by $A_{eff}=\int A(t)dt$, where

\begin{equation}\label{planeteq1}
A(t)=20(4-t \mathrm{v}_{s})+(R_{Jup}^{2}/2)(asin(10/R_{Jup})-10/R_{Jup}-\theta+sin(\theta))
\end{equation}
\begin{equation}\label{planeteq2}
cos(\theta)=(R_{Jup}-d_{0}+t \mathrm{v}_{s})/R_{Jup}.
\end{equation}

\begin{figure}[h]
   \begin{center} 
   \includegraphics[angle=0,width=7.in,clip]{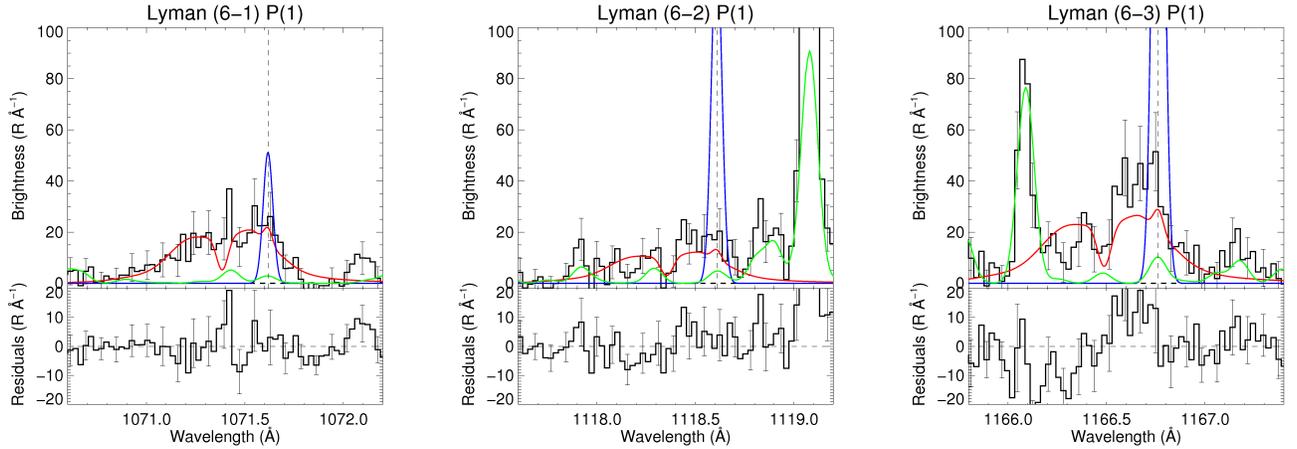}
   \end{center}
   \caption[$FUSE$ observations and fluorescence model for Jupiter. ]{ \label{jup} 
$FUSE$ MDRS Jupiter data, with PRD (red), CRD (blue), and electron impact (green) models overplotted. The residuals are shown in the bottom pannel, after subtracting both the PRD fluorescence model and the electron impact ecitation model (red and green).}
\end{figure}

\noindent The Jovian radius R$_{Jup}$ is 20\farcs34 (71,492~km), the scanning velocity of the slit v$_{s}$ is 0\farcs00383~s$^{-1}$, and the initial position of the leading edge of the slit, d$_{0}$, is 2\farcs68 from the edge of the disk. The integration results in 45\% slit coverage during the night time exposure. This scaling is employed in extracting the $FUSE$ spectra shown in Figure~\ref{jup}. Assuming that the limb starts approximately 0\farcs384 from the outer edge, we obtain that the 83\% of the recorded emission was originating in the disk, and 17\% in the limb. A similar method was used to compute the weights of each pencil beam integration to the final brightness predicted by the model, by replacing the planet radius by the position of the pencil beam as the limit for slit coverage in each case. This composite PRD model, shown by the red line in Figure~\ref{jup}, is in good agreement with the data, within the error bars. The blue line in the same figure represents the predictions of the CRD model, integrated in the same manner, while the green line shows the contribution from the electron impact model. Note that photon number conservation is satisfied by both CRD and PRD models. This comparison shows that the PRD model is a better description of the data, while the CRD approximation is clearly incompatible. Equivalently, the two models correspond to the limiting cases of $\gamma=1$ and $\gamma=0$ in Eq.~\ref{req4}, respectively. In addition to the energetic and collision rate arguments, this comparison between the CRD and PRD models and the data is further evidence that for the radiative transfer of the \htwo\ lines pumped by solar Ly$\beta$, collisional processes are negligible. Collisional scattering would reduce the amount of coherent re-emission in the line wings and redistribute more photons in the line core, contrary to observations. The remaining differences between the model and the data could be due to unaccounted for variations of the atmospheric composition across the limb of the planet. 

   \begin{figure}[h]
   \begin{center} 
   \includegraphics[angle=0,width=7.in,clip]{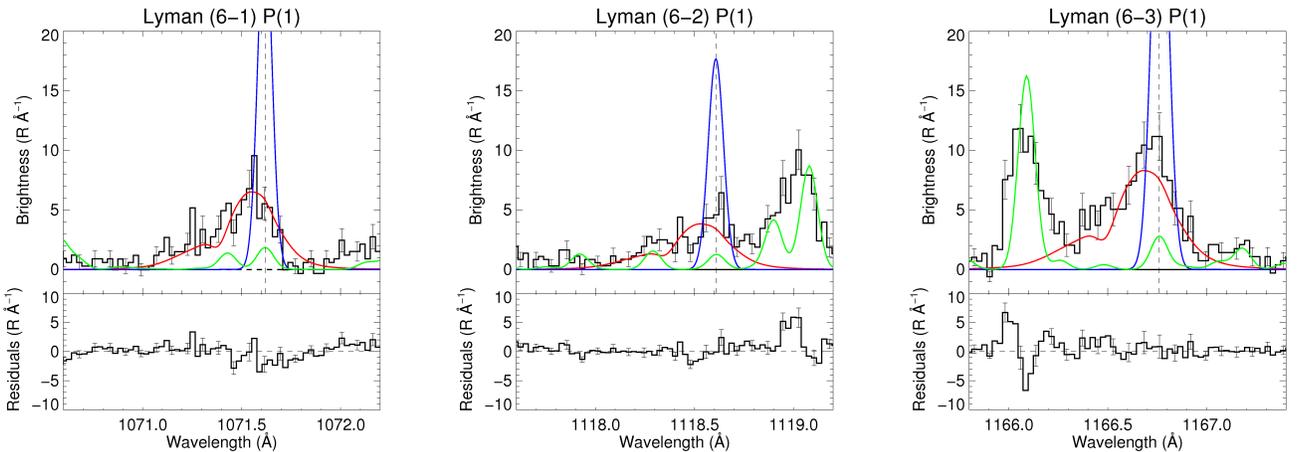}
   \end{center}
   \caption[$FUSE$ observations and fluorescence model for Saturn. ]{ \label{sat} 
$FUSE$ LWRS Saturn data, with PRD (red), CRD (blue), and electron impact (green) models overplotted. The residuals are shown in the bottom pannel, after subtracting the sum of the red and green curves.}
   \end{figure}

\subsubsection{Saturn}

As opposed to the the emission generated by electron impact excitation, restricted to the auroral region, the solar pumped fluorescence is integrated over the entire disk of the planet. The fluorescence models were calculated for a plane-parallel atmosphere, using the depth profiles for the gas density and temperature shown in the right panel of Figure~\ref{jupatm}. The atmospheric model was derived by \citet{Moses:2000}. A single pencil beam integration was sufficient to reproduce the $FUSE$ observations covering the entire disk of the planet. The PRD, CRD, and electron impact models are overplotted in red, blue, and green, respectively, on the $FUSE$ spectra in Figure~\ref{sat}. The Saturn data have a higher signal-to-noise than in the Jupiter case (used the 30\arcsec$\times$30\arcsec\ aperture), and we obtain a better agreement with the PRD model, while the CRD approximation is again inconsistent with the data.

\subsection{Preliminary Models for Continuum Pumped \htwo\ Fluorescence }{\label{h2nebmod}}

The occurrence of far-UV \htwo\ fluorescence in reflection nebulae can be confirmed by infrared observations of transitions from higher ro-vibrational levels ($v'>2$) in the ground electronic state, through which the fluorescent molecules cascade down to lower states. At expected gas temperatures below 1000~K, the LTE population of these hot ro-vibrational levels would be insufficient to produce the observed infrared line strengths. Such is the case for the infrared observations of the reflection nebulae NGC~2023 and NGC~7023 \citep{gatley87,Martini1999, Takami:2000}, which support the presence of significant far-UV \htwo\ fluorescence. Observations of these regions have been performed with $IUE$ and HUT longward of 1200~\AA\ \citep{francephd}, and with $FUSE$ at shorter wavelengths. Fluorescent \Htwo\ emission has been detected in the longer wavelength $IUE$ spectra, namely the features at 1575/1608~\AA, but not conclusively in the $FUSE$ data \citep{francephd}. The rocket observations of NGC~2023 \citep{Burgh:2002} also suggest a far-UV excess above the dust-scattered light, that might be due to molecular hydrogen fluorescence, but no individual transitions have been positively detected. We suggest that for the large optical depths characterizing these objects, the coherent scattering of the photons in the line wings is more important than previously thought, leading to a decrease of the line contrast. In this case, a high resolution spectrum can look qualitatively different than predicted by CRD models. Investigating this effect requires a model of the continuum-pumped \htwo\ fluorescence, including frequency cross-redistribution between all excited transitions ($>10^{3}$). Preliminary results show that the expected effects on the output spectrum are reproduced by the use of PRD.

   \begin{figure}[h]
\begin{center}
  \includegraphics[height=4.in]{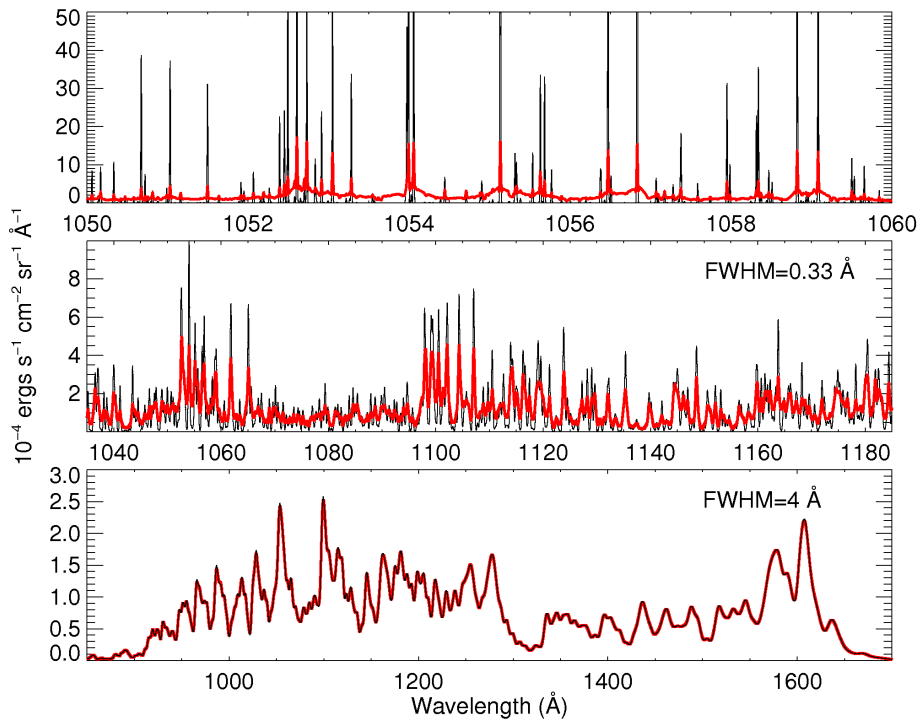}
  \end{center}
   \vspace{-0.2in}
  \caption[Comparison between the H$_{2}$ spectrum predicted by the CRD and PRD toy models for NGC 2023.]{Comparison between the H$_{2}$ spectrum predicted by the CRD (black) and PRD (red) toy models for NGC~2023. The top panel shows a region of the spectrum at the resolution of the numerical computation, while the middle and bottom panels show the same models convolved with 0.33~\AA\ and 4~\AA\ FWHM gaussian kernels, respectively.}
  \label{2023}
  \vspace{0.1in}
\end{figure}

We construct a toy model, with an incident 22,000~K blackbody spectrum normalized to 5000 times the Habing galactic mean \citep{Meyer:2001}, illuminating a uniform H$_{2}$ cloud with a rotational temperature of 1500~K, a Doppler $b$ parameter of 1.8~km~s$^{-1}$, and an integrated column of 5$\times$10$^{20}$~cm$^{-2}$. The radiative transfer was performed in a single layer approximation, to minimize computational resources. Excitation from higher ro-vibrational levels has not been taken into account. The results shown in Figure~\ref{2023} represent the first PRD calculation of a fluorescent molecular spectrum pumped by a continuum source. The upper panel shows a magnified section of the synthetic spectrum, with the PRD model overlaid in red on the CRD model. We note in the PRD model a decrease in the line contrast and a blending of the line wings, leading to the emergence of a quasi-continuum, where a significant fraction of the scattered photons is found. According to Figure~\ref{compprd}, we expect that the line peaks would be reduced even more when using a multi-layer approximation. The inclusion of dust scattering in the model would have a similar effect, effectively increasing the optical depth of the medium. The next panels show the same model convolved with gaussian kernels of 0.33~\AA\ and 4~\AA\ FWHM, respectively. These spectra correspond to the resolution of the $FUSE$ and $IUE$ observations, and demonstrate that at the resolving power of $FUSE$ there are differences between the two models that could be detected, while at the $IUE$ resolution the two models are indistinguishable, and we expect similar predictions for the 1575/1608~\AA\ features. In this sense, this model is qualitatively consistent with both the low signal strength of the fluorescent emission in the $FUSE$ data, and the emission features detected by $IUE$.

\section{SUMMARY}{\label{secsum}}

$FUSE$ observations of planetary atmospheres have revealed the need to include PRD in radiative transfer models for H$_{2}$ fluorescence. As shown in Section~\ref{genprd}, the inclusion of PRD will affect not only the integrated line strengths, and thus the escape probabilities, but also the estimated detectability of the lines due to the spread of the photons in the line wings. We constructed synthetic spectra for the atmospheres of Jupiter and Saturn, taking into account the depth dependence of the model parameters. The results are in very good agreement with the data, showing that the line profiles are reproduced only when accounting for PRD. Using an approximate integration method, we show that the use of PRD can significantly affect the resulting spectrum for the continuum pumped \htwo\ fluorescence, containing few$\times$10$^{3}$ overlapping transitions. The result is a decrease in the line contrast that can negatively influence the detectability of \htwo\ fluorescence with high resolution spectrographs.

The \Htwo\ lines pumped by solar Ly$\beta$ are very suitable for investigating the effects of PRD on the observed spectrum at high spectral resolution, due to the strong variation of the exciting spectrum over the absorbing line profile. Fluorescent molecular lines pumped by strong stellar features might also offer the possibility of detecting molecular hydrogen in extrasolar planets. In the most favorable circumstances, we find that the maximum ratio of such lines to the stellar continuum, assuming that the star and the planet are not spatially resolved, is only a few$\times$10$^{-4}$. Similar considerations should be applied to cases where multiple absorbing lines overlap over a narrow wavelength range, but observationally they are harder to disentangle. 
\acknowledgements

We thank the \fuse\ ground system personnel, particularly B.\ Roberts,
T.\ Ake, A.\ Berman, B.\ Gawne, and J.\ Andersen, for their efforts in
planning and executing these moving target observations. The limb-scan observations were designed by Alfred Vidal-Madjar and Lotfi Ben-Jaffel, members of the \fuse\ Solar System Guaranteed Time Program (P120). This work is partially based on data obtained for the Guaranteed Time Team by the
NASA-CNES-CSA \fuse\ mission operated by the Johns Hopkins University.
Financial support was provided by NASA grants NAS5-32985, NAG5-13085, NAG5-13719, and NNG04WC03G.

\appendix

This appendix presents the generalization of the source function used in Section~\ref{genprd} to include the collisional contributions. Taking into account that the transition rate out of level $j$ is $P_{j}=\displaystyle\sum_{k}(C_{jk}+R_{jk})$, we can write the population of level $j$ as
\begin{equation}\label{req3}
n_{j}=\frac{\displaystyle\sum_{i}n_{i}(C_{ij}+R_{ij})}{P_{j}}=\frac{\displaystyle\sum_{i}n_{i}C_{ij}}{P_{j}}+\frac{\displaystyle\sum_{i}n_{i}R_{ij}}{P_{j}}\equiv n_{j}^{C}+n_{j}^{R},
\end{equation}
where $C_{ij}$ and $R_{ij}$ are the rates of $inelastic$ collisions and radiative transitions, respectively.

 Substituting this expression in the line emission coefficient ( in Eq.~\ref{coeff}, or following the first sum sign in the numerator of Eq.~\ref{sfunc} ), we obtain
\begin{equation}
j_{ij}(\nu)=\frac{h\nu}{4\pi}A_{ji}D_{ij}(\nu),
\end{equation}
where we defined
\begin{equation}\label{deq}
D_{ij}(\nu)\equiv\phi_{ij}(\nu)\left[n_j-\gamma\displaystyle\sum_{k}\frac{n_{k}B_{kj}}{P_{j}}\int I(\nu')\phi_{kj}(\nu')d\nu'\right]+\gamma T_{ij}(\nu),
\end{equation}
and
\begin{equation}\label{teq}
T_{ij}(\nu)\equiv\displaystyle\sum_{k}\frac{n_{k}B_{kj}}{P_{j}}\int I(\nu')R_{kji}^{II}(\nu,\nu')d\nu',
\end{equation}
On the other hand, the population of level $j$ due to radiative transitions is
\begin{equation}\label{nreq}
n_{j}^{R}=\displaystyle\sum_{k}\frac{n_{k}B_{kj}}{P_{j}}\int I(\nu')\phi_{kj}(\nu')d\nu' = \int T_{ij}(\nu) d\nu,
\end{equation}
the last equality following from the normalization properties of the redistribution function, Eq.~\ref{eq6}. Substituting Eq.~\ref{nreq} and Eq.~\ref{req3} into Eq.~\ref{deq}, we obtain
\begin{equation}\label{deq2}
D_{ij}(\nu)=n_{j}^{C}\phi_{ij}(\nu)+(1-\gamma)n_{j}^{R}\phi_{ij}(\nu)+\gamma T_{ij}(\nu).
\end{equation}
This expression clearly separates all contribution to the line emission coefficient, and it is easy to implement, given that $n_{j}^{R}$ is easy to evaluate from Eq.~\ref{nreq}, when the quantities $T_{ij}(\nu)$ have already been computed. We see that in the absence of collisions ($C_{ij}$=~0 and $\gamma$=~1) $D_{ij}(\nu)$ reduces to $T_{ij}(\nu)$, and we recover the Eq.~\ref{req2}, after substituting the line emission coefficient into the source function.

\bibliography{ms_arxiv}

\end{document}